\documentclass[conference]{IEEEtran}
\IEEEoverridecommandlockouts

\usepackage{cite}
\usepackage{amsmath,amssymb,amsfonts}
\usepackage{algorithmic}
\usepackage{graphicx}
\usepackage{textcomp}
\usepackage{color}
\def\BibTeX{{\rm B\kern-.05em{\sc i\kern-.025em b}\kern-.08em
    T\kern-.1667em\lower.7ex\hbox{E}\kern-.125emX}}
\usepackage{bm}
\usepackage{graphicx}
\usepackage{subcaption}
\begin{document}

\title{Distributed Data Vending on Blockchain}


\author{\IEEEauthorblockN{Jiayu Zhou$^1$, Fengyi Tang$^1$, He Zhu$^2$, Ning Nan$^2$, Ziheng Zhou$^3$}
\IEEEauthorblockA{
$^1$Computer Science and Engineering, Michigan State University, East Lansing, MI 48824, USA\\
$^2$BitOcean Ltd., Tokyo, Japan\\
$^3$Vechain Foundation, Singapore\\
Email: \{jiayuz, tangfeng\}@msu.edu, \{hezhu, nanxiaoning\}@bitocean.com, peter.zhou@vechain.com}}


\maketitle

\begin{abstract}
Recent advances in blockchain technologies have provided 
exciting opportunities for decentralized applications. 
Specifically, blockchain-based smart contracts enable 
credible transactions without authorized third parties. 
The attractive properties of smart contracts facilitate \emph{distributed data 
vending}, allowing for proprietary data to be securely exchanged on a 
blockchain. Distributed data vending can transform domains such as healthcare 
by encouraging data distribution from owners and enabling large-scale data aggregation.  
However, one key challenge in distributed data vending is 
the trade-off dilemma between the effectiveness of data retrieval, and the 
leakage risk from indexing the data. In this paper, we propose 
a framework for distributed data vending through a combination 
of data embedding and similarity learning. We illustrate 
our framework through a practical scenario of distributing and aggregating 
electronic medical records on a blockchain. Extensive empirical 
results demonstrate the effectiveness of our framework. 
\end{abstract}


\section{Introduction}

Recent years have witnessed surging interests in blockchain
technology~\cite{tschorsch2016bitcoin}, an open and decentralized ledger that
links transaction records using blocks and provides proven security backed by
cryptography. Blockchain was first introduced as the infrastructure
supporting the electronic cash system Bitcoin~\cite{nakamoto2008bitcoin}, and
is now transforming a wide domain of industries such as supply
chains~\cite{vechain2018web}, biomedical research~\cite{kuo2018modelchain},
healthcare~\cite{ekblaw2016medrec,peterson2016blockchain,kuo2017blockchain},
financial transactions~\cite{iansiti2017truth},
networking~\cite{zhu2018edgechain} and social
networks~\cite{iansiti2017truth}. Of particular interest is the Ethereum
blockchain~\cite{wood2014ethereum}, which has enabled credible transactions without
trusted third parties by introducing \emph{smart contracts}, with small programs
running on top of the blockchain to provide transparent and secured contracting
with reduced transaction costs.

The advent of blockchain infrastructures paved the way for a new domain of
\emph{data vending}, allowing individual data owners to directly benefit from
sharing proprietary data over the blockchain. In the era of big data, vast
amount of data collected has been widely used to improve decision making for
industries, namely through building personalized recommender
systems~\cite{linden2003amazon,koren2009matrix} and targeted
advertisement~\cite{cui2005content}. 
As a result, organizations that collect and aggregate data at scale 
stand to profit enormously in the process. 
However, as data stakeholders, the users from whom the data
is collected from rarely get their share of dividends despite significant
contributions to the fortune. In fact, in most cases, organizations
 regard collected data as their private assets and prevent the data from
being shared even for research purposes, which could otherwise contribute to the advancement of our society.

Perhaps a prime example lies in our healthcare system, where electronic health records (EHR)
systems are now deployed in most hospitals across the United States. In recent years,
medical histories in EHR have been used to build data-driven models to improve
healthcare resource management~\cite{zhou2013feafiner,zhou2014micro}.
 Analyses of the EHR data often reveal important
insights into the underlying pathophysiology of numerous complicated diseases~\cite{zhou2013modeling}. These discoveries are invaluable to the development of drugs and
treatments. However, the patients who own the data~\cite{esposito2014health} and contribute to research outcomes are rarely
rewarded for their contributions, despite the hiking healthcare
costs which accompany the rise of healthcare information exchange in recent years~\cite{greene2017targeting,livingston2018health}. 
On the other hand, many
research institutions such as universities have an extremely difficult time to
access existing health data because of strict regulations such as HIPPA~\cite{us2003hippa} and
bureaucracy, even though there are individuals willing to share their medical
histories, especially when they are incentivized to do so.

\begin{figure}[t!]
\includegraphics[width=0.5\textwidth]{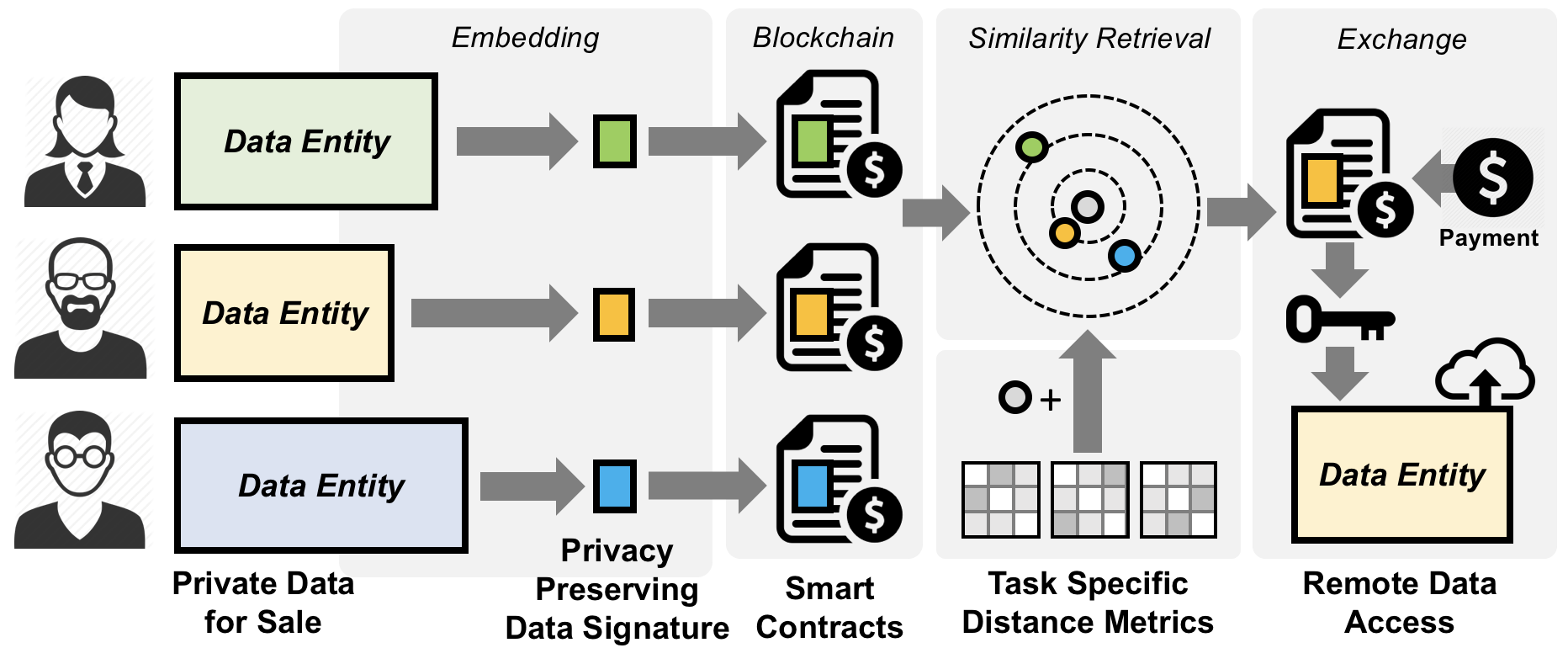}
\caption{Illustration of the proposed distributed data vending (DDV) framework. 
DDV enables private data to be exchanged through smart contracts. }
\vspace{-0.2in}
\end{figure}

In this paper, we propose a \emph{distributed data vending (DDV)} 
framework which consists of a suite of machine learning modules to
enable individuals to exchange data on blockchains through smart contracts. Like
other vending services, a data entry has to be indexed and retrieved before it
can be exchanged on blockchains. However, the challenge comes from the dilemma
between data indexing and information leakage: building indexes for retrieval
requires access to the data, which is hard to achieve on blockchains where no
trusted third parties are available. To tackle this challenge, the proposed
DDV approach generates a signature of the data entry using a \emph{privacy-preserving}
data embedding procedure. The signature can be published along with the
data access in smart contracts. The framework also provides task-specific
similarity functions, which take the signatures as input to measure the
similarity among data entries. The proposed DDV framework can be used in many
data vending scenarios, and in this paper we use EHR data vending as a
concrete example to illustrate the concept and its feasibility. We also provide
extensive empirical evaluations to demonstrate the effectiveness of the
proposed framework.

\section{Related Work} 

Even though the distributed data vending is a rather new problem, the
challenges of this problem are closely related to many existing studies, such
as health data management based on blockchain, data embedding techniques and
distance metric learning in machine learning. We will briefly survey these 
areas and point out the relationship to this study. 

\subsection{Blockchain Applications on Healthcare Data}
The properties of blockchain make it a promising tool in many health
informatics applications~\cite{kuo2017blockchain}: from building decentralized
backbones for health data exchange and interoperability, protocols enforced by
immutable ledgers that keep track of clinical research~\cite{topol2016money},
data provenance and robustness of medicine production through pharmaceutical
supply chain~\cite{baxendale2016can,livingston2016health}, to maintain patient
privacy and EHR security through the persistence of consent statements in
blockchain~\cite{brodersen2016applying}. The proposed DDV framework utilizes
an existing blockchain infrastructure to facilitate data exchange, and
provides an effective tool to collect medical data and thus accelerate medical
research.
 
\subsection{Medical Information Exchange on Blockchain}
Of particular interests is the domain of medical information exchange, where
many studies have been done to facilitate the second use of EHR data for
clinical/biomedical research. Examples include
MedRec~\cite{azaria2016medrec,ekblaw2016medrec},
Healthbank~\cite{mettler2016blockchain} and
ModelChain~\cite{kuo2018modelchain}.
MedRec~\cite{azaria2016medrec,ekblaw2016medrec} is closely related to the
proposed research. It proposed a blockchain-based data management system to
consolidate the fragmented medical records. The system involves patients in
the data life cycle, allowing the patients to take control of the permission
of their own medical records, and encouraging medical stakeholders to
participating the blockchain network with incentives of aggregated data. The
work delivered an important message that patients are one of the data owners,
can serve an important role in data sharing among stakeholders, and should be
controlling the data sharing process. However, one major issue in this
framework is that the data is persistent in a data center of the hospital
system, which requires the API exposure for remote access. Such practice is
usually hard to implement due to the excessive security risks of external
APIs. Moreover, since query strings of the database are persistent in smart
contract, it would incur significant maintenance costs during data
migration~\cite{velimeneti2016data}. In the proposed DDV framework, we make no
assumptions on the physical location of data. Once the smart contracts
including desired information are retrieved, the contracts can grant access to
data stored anywhere in the cloud. 

\subsection{Data Embedding and Distance Metric Learning}
In machine learning research, data embedding has been studied for decades. The
embedding can be learned in a supervised fashion (e.g., linear discriminant
analysis~\cite{mika1999fisher}) which learns a low-rank subspace that captures
most of the discriminative information, or an unsupervised one (e.g.,
principal component analysis~\cite{jolliffe1986principal}), which learns a
subspace to capture the geometric manifold of the data matrix. More recently, 
the advances of deep learning lead to highly non-linear embedding techniques such 
as autoencoders~\cite{vincent2008extracting} as well as recurrent neural networks 
for embedding time sequence data~\cite{sutskever2014sequence}. 
In many information retrieval systems we need to provide a distance metric to
evaluate the similarity between pairs of data points, and distance metrics
vary from task to task. The distance metric learning~\cite{yang2006distance}
is one type of supervised learning that learns a metric from data for one
specific retrieval task. 
In medical informatics, for example, the study of patient similarity attracts
much research efforts, which facilitates effective patient stratification and
discovers important risk factors~\cite{brown2016patient}.

In this paper, we propose to use a two-stage approach to combine the data embedding 
and distance metric learning, which firstly learns a privacy-preserving 
signature to effectively summarize the original private data, which is then 
used to compute similarity by a chosen similarity metric learned from database, 
to achieve retrieval of smart contracts, without compromising the integrity of 
the private data.

\section{Privacy-Preserving Data Vending}

Data vending is the exchange of private data between individual \emph{data
providers} (owners) and \emph{data consumers}. The purpose of distributed data
vending (DDV) is to enable data providers to use existing blockchains as
infrastructure to list the data. \emph{Data consumers} then retrieve data from
blockchain and complete the purchase. The entire data exchange process is done
without trusted third parties involved. In this section, we introduce the DDV
framework and its key components. We use vending of medical records as a
running example in the remaining of this paper and we note that the same
technique can be extended to the setting of general data vending\footnote{By
general data vending we mean the data is structured and can be represented in
a vector/matrix/tensor form}.

\subsection{Distributed Data Vending}

Assume a data provider ($i$) would like to sell some private data to others
for profits and the data entity can be represented in the form of a multi-dimensional
tensor $\mathcal X_i \in \mathbb R^{d_1 \times \dots \times d_k}$ (which subsumes the matrix case $X_i$ and the vector
case $\mathbf x_i$), where $d_1  \dots  d_k$ are the dimensionality of $k$ modes. For example, the purchasing history of a person can be represented in a tensor capturing 
[\emph{item} $\times $ \emph{time} $\times $ \emph{location}], or the watching history for a set of users using [\emph{item} $\times $ \emph{user} $\times $ \emph{time}] tensor~\cite{karatzoglou2010multiverse}. And the historical medical records of a patient can be represented in the form of a sparse matrix~\cite{zhou2014micro} capturing [\emph{time} $\times $ \emph{diagnosis}].

In order for potential data consumers to find the
availability of $\mathcal X_i$, the data provider needs to list the
information of the data to a listing service, such that the consumers can
query the listing by sending a relevance criteria. The payment and data 
exchange can be done once the data entity is identified as relevant and fits the budget of the consumer. 

Even though we are using many of such listings everyday, for example Amazon for
general merchandise and Netflix for movies, a listing for general data vending
is very different. Merchandise can be listed according to categories and
properties, where as a movie or a song can be retrieved from its artist, genre
and the year of release. However, there is no predefined categorical
information for an arbitrary data tensor $\mathcal X_i$, and thus it is  
challenging to retrieve the data tensor needed. 
When vending medical data, one potential data
consumer may be interested in knowing answers to query questions such as
``\emph{did this patient have any diagnosis of hypertension before?}'' or
``\emph{how many past encounters of renal failures in the past three
years?}'', before deciding whether or not to purchase the data. In order to
provide response to such query, the listing service has to know all the
details of the data, i.e., the listing service has access to the entire data
to be exchanged.

When the listing service is a trusted 3rd party, such query is feasible.
However, such listing service is associated to an extremely high security
risk: once the 3rd party listing server is compromised, all the data entries
on the listing server will be leaked. Different from other listing services,
once the data themselves are leaked there is no way to control the perimeter
of damage. As such, there is a strong demand of a distributed data vending
framework, facilitating a decentralized vending procedure that does not
involve the listing service from a trusted 3rd party. \emph{Smart contracts}
has provided a secured and efficient tool for vending contracts on top of
blockchains. However, without the 3rd party listing, the retrieval problem now
involves a dilemma: On one hand, we want to enable effective retrieval for 
data consumers to identify the data entities of interest; on the other hand, 
indexing data entities requires detailed information of the data, which can 
leak the content of data entities before they are purchased for use. 

To illustrate this leakage in the context of medical records, assume that we
expose the number of diagnoses of \emph{Congestive Heart Failure (CHF)} and
\emph{Chronic Kidney Disease (CKD)} in the smart contract to enable indexing.
Any data consumer now can simply iterate through all available smart contracts
in the blockchain, and the consumer can explicitly compute the marginal
distributions $\Pr(\text{CKD})$ and $\Pr(\text{CHF})$, the joint distribution
$\Pr(\text{CKD}, \text{CHF})$, as well as conditional distribution of
$\Pr(\text{CKD}| \text{CHF})$ and $\Pr(\text{CKD}|\text{CHF})$. The data
leakage does not stop at statistics level, as such distributions allow one to
directly build predictive models, say, using Na\"{i}ve Bayes, to infer
posterior probabilities of any diseases used for indexing. However, to
acknowledge the data contribution to such predictive model, any data provided
involved should be paid through executing the associated smart contracts. 

In order to balance retrieval efficiency and data leakage without a 3rd party
listing, we propose a distributed data vending (DDV) framework. In this
framework we first embed the data into a private preserving signature vector,
and then maintain a set of public similarity metrics. The signature of a data
entity are publicly available in smart contracts selling the entity, and the
retrieval is done by combining the data signatures and similarity metrics.

\begin{figure}
\centering
\includegraphics[width=0.42\textwidth]{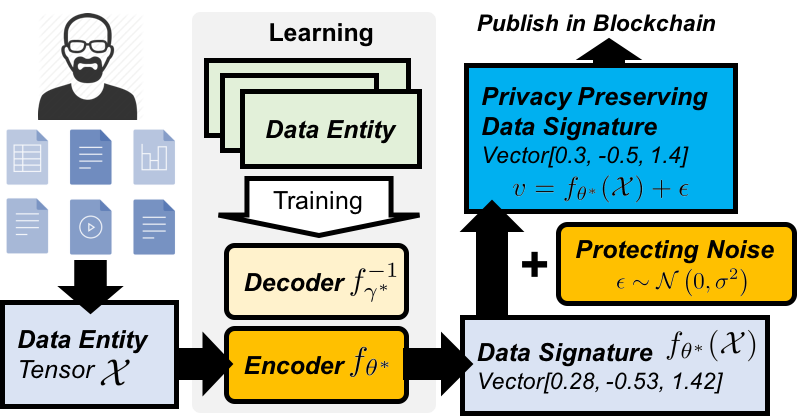}
\vspace{-0.05in}
\caption{Illustration of the learning of privacy-preserving signatures in
DDV.}
\label{fig:train_signature}
\vspace{-0.22in}
\end{figure}

\subsection{Data Signature}

We propose to use data embedding techniques to compute data signature. 
The goal of data embedding is to project a data entry to a usually lower-dimensional subspace such that the data entry can be represented by a low-dimensional vector. Formally, we denote an embedding function $f$ that projects a
data tensor $\mathcal X_i \in \mathbb R^{d_1 \times \dots \times d_k}$ to a vector $z_i \in
\mathbb R^p$, and thus $f_\theta (\mathcal X_i) \rightarrow z_i$, where the vector $\theta$ 
includes a set of parameters of the projection function. One way to obtain the
vector representation to extract a set of summary statistics from the tensor,
and however, it is hard to warrant the effectiveness of retrieval because of
the difficulty from estimating how much information are covered by summary
statistics. On the other hand, different data entities may have different
shapes/dimensionality, and it is challenging to design such statistics
manually. 

When there are many data entities available, the data embedding can
be done in a data-driven way, i.e., we learn a projection function such that
the subspace learned can be maximally recover the data entities. Let $\mathcal
D = \{\mathcal X_1, \mathcal X_2 \dots, \mathcal X_n\}$ be a set of $n$ data
entities available for training, the data-driven embedding is given by the 
following objective:
\begin{align}
\min_{\theta, \gamma}  \sum\nolimits_{i=1}^n 
   \ell\left( \mathcal X_i,  f_{\gamma}^{-1} (f_{\theta} (\mathcal X_i)) \right), \label{eqt:embedding_obj}
\end{align}
where $\ell (\cdot,\cdot)$ is a loss function that evaluate the recovery
error, and $f_{\gamma}^{-1}$ is an inversion function of $f$ that tries to
reconstruct the data entity from the embedded vector, and $\gamma$ is the 
parameter of the inversion function. $f$ is usually called \emph{encoder} 
and $f^{-1}$ is corresponding \emph{decoder}. 

Let $\theta^*, \gamma^*$ be an optimal solution pair to
\eqref{eqt:embedding_obj}, then we can publish $\theta^*$ and the functional 
form of $f$, for example to a blockchain. Then any data provider who wants to
sell data $\mathcal X_j$ can now compute the embedding $v_j =
f_{\theta^*}(\mathcal X_j)$ and publish $v_j$ along with the smart contract.
We call this embedding vector the \emph{signature} of the data matrix $\mathcal X_j$.
Details of this publishing will be illustrated in
Sec.~\ref{sec:method:contract}. Next we discuss two important issues
associated to the embedding process.

\paragraph{Incremental Signature Learning}
A high performance projection $f_{\theta}$ guarantees that 
the signature includes a comprehensive description of the data entry been 
projected. Therefore learning a good projection function is the key  
for effective retrieval. Since the projection function depends on 
the data entries $\mathcal D$ used for training, and usually we do not 
have a large amount of data to start. As more data entities are available, 
the projection function can be updated to improve its performance. This 
suggests an online learning approach to incremental update parameters of
the embedding function $\theta$. For example, using:
\begin{align*}
[\theta^+, \gamma^+]  = [\theta^-, \gamma^- ] - \tau \nabla_{[\theta, \gamma]}   
\ell\left( \mathcal X_i,  f_{\gamma^-}^{-1} (f_{\theta^-} (\mathcal X_i)) \right),
\end{align*}  
where $\tau$ is the step size for incremental update, which is converging to
0, and the superscripts $+$ and $-$ denote before and after the update
respectively.

\paragraph{Privacy-Preserving Signature}
The rationale behind releasing the signature $v_j$ to public is the belief
that $v_j$ cannot be used to recover the original data entry $X_j$.
Conceptually this is highly likely because that the original data $X_j$ is in
a much larger space than the signature $v_j$, and such projections are
supposed to single directional, and given $v_j$ the chance of recovering the
original space is small. However, because that the embedding projection is
learned in a way that seeks an structured manifold of the original data in
$\mathcal D$, it is possible to recover the original data once the shape of
the manifold is obtained. And manifold information is captured in the inverse
function $f_{\gamma}^{-1}$, which is why the reconstruction $f_{\gamma^*}^{-1}
(f_{\theta^*} (\mathcal X_i))$ can be quite small in practice. Therefore
protecting $f_{\gamma^*}^{-1}$ is essential to protect the signatures. Once an
accurate approximation of $f_{\gamma^*}^{-1}$ is obtained by adversarial
parties, they can use it to recover a large amount of information by going
through the signatures alone.

On the other hand, since $f$ is publicly available, it is reasonable to assume
that the functional form of $f^{-1}$ is also publicly available. This implies
that once the parameters in $\gamma$ are estimated, then the signatures are
compromised. In fact, when someone already purchased quite a few data entries, 
the purchased data entries can be used to construct a dataset $\tilde{\mathcal D}$, 
which can then be used to estimate the decoder:
\begin{align}
\tilde \gamma = \text{argmin}_{\gamma}  \sum\nolimits_{\mathcal X \in \tilde{\mathcal D}} 
   \ell\left( \mathcal X_i,  f_{\gamma}^{-1} (f_{\theta^*} (\mathcal X_i)) \right), \label{eqt:attack}
\end{align}
where $\tilde \gamma$ is the estimated decoder from $\tilde{\mathcal D}$.
Therefore we propose to apply a Gaussian noise to the computed signature to
destroy the manifold structure, i.e., given a data entry $\mathcal X$, we 
compute the \emph{privacy-preserving signature} $v$ by:
\begin{align}
v = f_{\theta^*} (\mathcal X) + \epsilon, \quad \epsilon \sim \mathcal N \left(0, \sigma^2\right), \label{eqt:signature}
\end{align}
where $\sigma^2$ is the variance controlling the magnitude of the noise, and
should be chosen to balance the retrieval effectiveness and privacy level: a
higher magnitude of noise gives stronger privacy protection but weakens the
retrieval performance, and vice versa. We illustrate such trade-off in the
empirical study section. The procedure of obtaining the signature is
summarized in Figure~\ref{fig:train_signature}.

\subsection{Similarity Retrieval}
Once a privacy-preserving signature $v$ is computed using
\eqref{eqt:signature}, a data provider can then create a smart contract and
publish the signature in the smart contract. Since the blockchain is publicly
available, and data consumers can now get the list of signatures for all
available data entries. In order to retrieve a set of relevant data entries to
purchase, a consumer can complete the retrieval by comparing the signature of
the purchase with a query signature $\hat v$. Directly comparing two signature
vectors through cosine similarity $\cos(v, \hat v)$ or inner product $v^T \hat
v$ may not be effective, as the subspace of $v$ is chosen to maximize the
reconstruction and thus may not be specific for the desired retrieval task.
For example, one consumer wants to retrieval medical records that 
have prior diabetes diagnoses, and another consumer is interested 
in building models from those identified as dementia. For these two  
retrieval tasks, given the same query signature $\hat v$, the ranked 
list of data entities should look very different, because that the 
the definitions of ``similarity'' are different. 
From the machine learning perspective, the subspace learning for signature
belongs to the unsupervised learning paradigm, and we need a supervised
learning procedure and learn a similarity function $s_{t}(v,
\hat v)$ that is specific to task $t$, to carry out an efficient retrieval.
Equivalently, we can learn a distance function $d_{t}(v, \hat v)$ for the same
purpose (by reversing the retrieved rank).

Let the $d_{t}$ be the Mahalanobis distance, i.e., $d_{t}(v, \hat v) = \sqrt{(v -
\hat v)^T M_t (v - \hat v)}$, where $M_t$ is the parameter matrix that instructs  
how the distance should be computed for a specific task. Given the dataset
$\mathcal D_t$ for retrieval task $t$, the parameter matrix $M_t$ can be
estimated using a distance metric learning loss, e.g., the 
large-margin~\cite{parameswaran2010large}. We denote the loss function as
$\ell_\varepsilon$. When there are many retrieval tasks, their parameters can
be estimated jointly using the multi-task learning paradigm, to transfer
knowledge among tasks~\cite{parameswaran2010large,zhou2011malsar}:
\begin{align*}
\min_{{M_1, \dots, M_t }} 
\sum\nolimits_{t=1}^T & \left( \sum\nolimits_{(i, j) \in \mathcal P_t} d_{t} (v_i, v_j)^2 + \right. \\
  &\left. \sum\nolimits_{(i, j, k )\in \mathcal I_t} \ell_\varepsilon\left(d_{t}(v_i, v_j), d_{t}(v_j, v_k)\right)\right)
 ,\\
 s.t. \ M_t \succeq 0, \forall t, & \quad [M_1, \dots, M_T] \in \mathcal M,
\end{align*} 
where $d_{t}$ is parameterized by $M_t$, $\mathcal P_t$ is the partial order
set in dataset $\mathcal D_t$, and $\mathcal I_t$ is the triplet order set in
$\mathcal D_t$ and each triple $(i, j, k)$ indicates the distance between
$d_{t}(v_i, v_j)$ is closer than $d_{t}(v_i, v_k)$, and the positive semi-
definiteness of $M_t$ makes the objective a convex programming. The subspace
$\mathcal M$ constrains the solution spaces of $M_t$ and connects them to
enable knowledge transfer. The subspace structure can be defined using many of
existing multi-task learning approaches~\cite{zhou2011malsar}, such as shared
subspace basis, sparsity patterns, or composite structures. 

The multi-task metric learning approach can effectively improve the
performance of distance metrics when there are limited amount of data
supervision available, which improves the generalization of metrics through
knowledge transfer. The simpler single task learning approach can be used once
a large amount of training data is available, which reduces the additional
bias introduced by multi-task learning priors. When distance metrics are
learned for retrieval tasks, the metrics can be published online to a publicly
available metric library, with the descriptions of tasks that they are
associated to.

\subsection{Distributed Data Vending Workflow}
\label{sec:method:contract}

\begin{figure}[t!]
\includegraphics[width=0.48\textwidth]{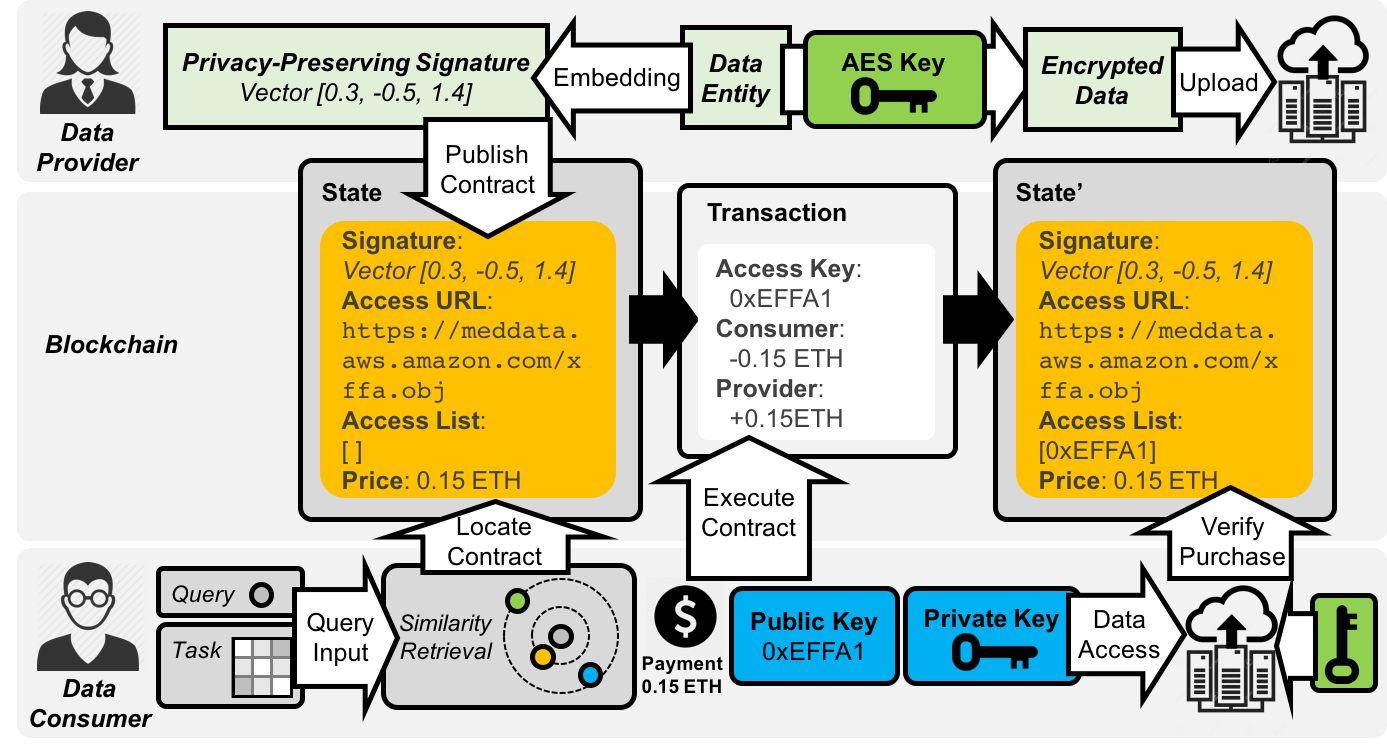}
\vspace{-0.05in}
\caption{Illustration of workflow for the proposed distributed data vending. }
\label{fig:workflow}
\vspace{-0.2in}
\end{figure}

In this section we summarize the workflow of the proposed DDV framework. We
assume the public availability of the following building blocks: the encoder
$f_{\theta^*}$ and a metric library that includes a set of well-trained
distance metrics. These metrics can be updated over time to improve the 
effectiveness of the retrieval. 

\paragraph{Data Provider} A data provider is an entity who owns private data and
can potentially sell the data pieces on blockchain.

\paragraph{Provider Device} A provider device is owned and trusted by a
\emph{data provider}. A typical provider device can be a smartphone. When a
\emph{data provider} wants to sell a piece of data, the raw data is firstly
formatted into a data entity, representing the data into a tensor (or
matrix/vector). The encoder $f_{\theta^*}$ is then used to embed the data into a
privacy-preserving signature vector. After that the provider device generates an
AES key and uploads the data encrypted by this key to a \emph{data server}. The
\emph{data provider} then creates a smart contract in the blockchain as
described above.

A purchase from any consumer triggers a decryption process where the
\emph{data provider} sends the AES key to the \emph{data consumer} via a secure
channel after handshaking.  The \emph{provider device} then regenerates an AES key,
encrypts the data using the new AES key again and sends the new copy of encrypted
data to the \emph{data server} for the next purchase.

\paragraph{Data Server} A data server hosts the encrypted data entities from
\emph{data providers}. Whenever a download request is initiated by a \emph{data
consumer} on a certain data entity, the server first retrieves its associated
smart contract in blockchain. When the \emph{data consumer} is authenticated by
its public key in the smart contract, the encrypted data can be be downloaded
by the \emph{data consumer}. The data server then requests a new encrypted data
for the next download request. This data ensures that each time a download is
verified, a new AES key will be used for the next download. This way each
encrypted copy of data entity will be valid for only once, so that the data will
not be at risk if the data server gets compromised. See remarks below for how the AES
key is deployed and retrieved.

\paragraph{Blockchain} The blockchain infrastructure can be any chain that
enables smart contracts, such as Ethereum~\cite{wood2014ethereum} or
VeChain~\cite{vechain2018web}. Whenever a data provider lists a data entity for
sale, a smart contract is created that includes the data signature, an access
URL of the data server (e.g., an AWS server) or an API address for retrieval, a
list of public keys that granted data access, as well as  the selling price for
the data access. The smart contract can also include many more details such as
demographic information of the patient if the data is EHR. 

\paragraph{Data Consumer} A data consumer starts with a data collection task in
mind, uses an existing data for query or construct the query, and applies the
embedding function to compute the query signature. Based on the retrieval task,
the provider then obtains the task-specific distance metric to rank the
similarity of data entries available for purchase. Once identified one (or more)
smart contract that includes data for purchase, the consumer then provides the public
key and the demanded payment to the blockchain to execute the contract. Once the
transaction is confirmed in the blockchain, the consumer then authenticates itself using
its private key to download encrypted data from the \emph{data server}. Meanwhile,
the data consumer retrieves the AES key from \emph{provider device} via the established
secure channel between them. The download begins once the payment is verified by the
server and the data is decrypted using the AES key.

The workflow is summarized in Figure~\ref{fig:workflow}. In the next section we
show a concrete example how the proposed DDV can be used for EHR data vending.

\noindent\emph{Remarks on Authentication and Data Security:} The
AES key is generated by and stored on the \emph{provider device}, while the
encrypted data entity is stored on the \emph{data server}. The isolation of the
key and the encrypted data prevents unauthorized decryption in case a \emph{data
server} is compromised. Secure channels between \emph{provider devices} and
\emph{data consumers} are established by handshaking using public/private keys of
both parties. The distribution of AES keys are done via the secure channels. The
one-time AES key and encrypted data entity have further enhanced the data security
of the framework.

\section{Case Study: Vending Healthcare Records}
\label{sec:case_study:medical_records}

\begin{figure*}
\centering
\includegraphics[width=0.9\textwidth]{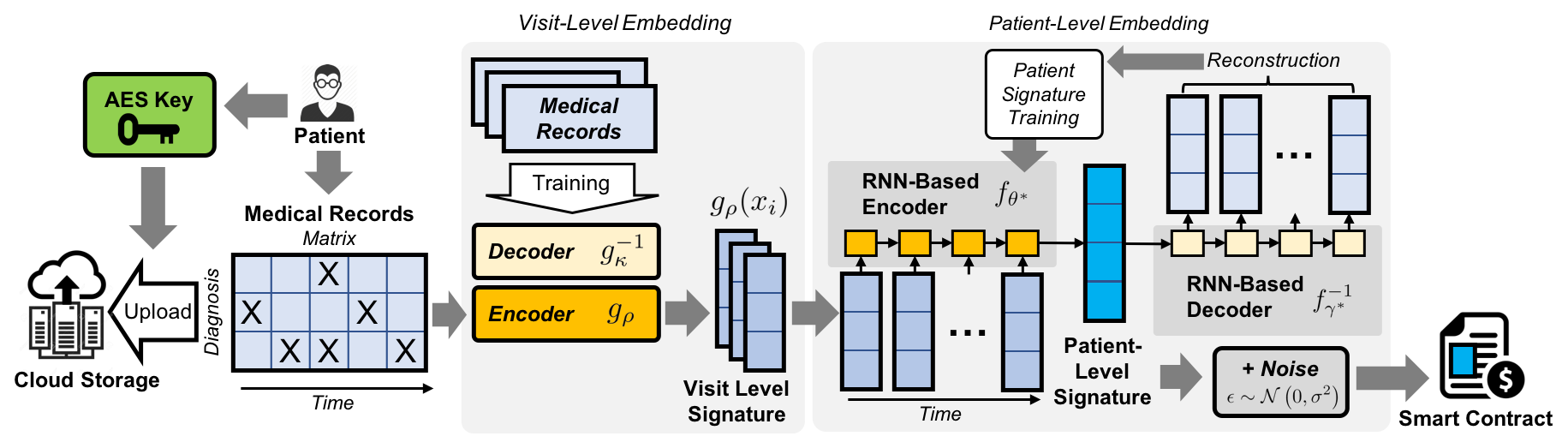}
\vspace{-0.1in}
\caption{Generating patient signature for vending medical records using the proposed DDV framework. }
\label{fig:vending_EMR}
\vspace{-0.2in}
\end{figure*}

The limited access to medical data has been a major bottleneck to many medical
studies. With more data available, clinicians and data scientists can develop
more accurate disease models and drug effects to reduce health costs and save
lives. We believe that the proposed DDV framework can be used to relieve the
the scarcity of data in medical research by encouraging patients to share
self-reported medical history through blockchain, and the data contribution is
acknowledged and rewarded by monetary incentives. In this section, we
illustrate a prototype implementation of our DDV framework in the context of
vending healthcare records. The implementation includes two components: (1)
incrementally learns signature representations of longitudinal patient
information and (2) produces task-specific similarity metric for data
retrieval.


Although previous studies have cited the effectiveness of end-to-end sequence-
to-sequence autoencoders over a wide array of representation
tasks~\cite{luong2015multi}, substantial data is often required to obtain
high-level performance using this approach. In the EHR setting, medical
records often vary in sample size depending on the granularity of data
available. For example, we have around $20$ times more visit-level data than
patient-level temporal data in our experimental database. This means that a
high-performing visit-level signature can be trained before sufficient data is
available to drive high-performing patient-level signatures, which require
many samples to capture both temporal patterns at the patient-level as well as
sparsity patterns at the visit-level. Thus, we decompose the signature
learning tasks into two sub-tasks, visit- and patient-level signatures, so
that our representation framework is better suited for incremental online
learning.

\paragraph{Visit-Level Signatures}
Obtaining the visit-level signature is a dimensional reduction task which
takes visit vectors $x_i \in \mathcal X$ and outputs embedded vectors
$g_{\rho}(x_i)$, where $g_{\rho}(.)$ is the visit level signature function
parameterized by $\rho$. In this case, we do not consider differences between
patients and simply try to learn a low-dimensional representation of medical
visits. For each visit, we consider all the relevant diagnostic ICD-9 group
codes. Assume there are $d$ ICD-9 diagnosis group codes for consideration, and
training inputs take on $x_i \in \mathbb{R}^{d}$ one-hot vector format, with
$1$'s denoting group codes present for the visit.
We then project the training vector into a $q$-dimensional dense latent space
$g_{\rho}(x_i) \in \mathbb{R}^{q}$, and $q\ll d$. 

We formulate the representation process as a multi-label classification problem, with $d$ binary classifiers trained in parallel for each visit sample. 
We learn the pair of encoder $g_{\rho}$ with the help of a decoder $g^{-1}_{\kappa}$ parameterized
by $\kappa$. For a vector $x$, let $\hat x^{\rho, \kappa}_i =
g^{-1}_{\kappa}(g_{\rho}(x_{i}))$ be the reconstructed vector. The parameters
$\rho, \kappa$ are then estimated from minimizing the following reconstruction
error:
\begin{align*}
\min_{\rho, \kappa} \sum_{j=1}^{d} \lambda_{j} 
   \sum_{x_{i} \in \mathcal X} \left\{
         x_{i,j} \log\left(\hat x_{i,j}^{\rho, \kappa}\right) 
         + (1-x_{i,j}) \log\left(1- \hat x_{i,j}^{\rho, \kappa}\right) \right\},
\end{align*}
where $\lambda_j$ is the weight that is given to the particular task $j\in [1,
d]$ which correspond to the group code to be predicted during recovery.
Depending on the sample, $\lambda_j$ assumes $1.0$ if the sample is negative
for the group code and $3.0$ if the sample is positive for the group code
during that visit. We used multi-layer perceptron to implement both $g_{\rho}$ and $g^{-1}_{\kappa}$.

\paragraph{Patient-Level Signatures}
Once visit-level signature embedding is learned, the original EHR tensor
$\mathcal{X}$ can be transformed into $\mathcal{G}$, which contains embedded
visits for each patient across time. This way, when a vectorized
representation of sequential patient information, we by-pass the sparse
recovery problem mentioned in the previous section. Therefore, in
\eqref{eqt:embedding_obj} we simply minimize over the mean-squared error (MSE)
between the original input $\mathcal{G}$ and the recovered tensor
$f^{-1}(\mathcal{G})$ from the embedding process:
\begin{align*}
\ell\left({G}_i, f_\gamma^{-1}(f_\theta(\mathcal{G}_i))\right) &= \left\|\mathcal{G}_i - f_\gamma^{-1}(f_\theta(\mathcal{G}_i))\right\|^2_F,
\end{align*}
where $f_\theta$ and $f_\gamma^{-1}$ are two recurrent neural networks (RNN). 
Specifically, we applied long-short term memory (LSTM) units to construct 
RNNs for the encoder and decoder. 
We note that the number of visits differ for patients, so the number of visit
vectors recovered per patient depends on the input patient. Thus, the RNN 
vector representation of $\mathcal{G}$ is robust to variations in visit
frequency of patients. 
The entire procedure for generating patient signature is summarized in Figure~\ref{fig:vending_EMR}.



\section{Experiments} 

In this section, we demonstrate the proposed distributed data vending (DDV)
framework using medical records as elaborated in
Section~\ref{sec:case_study:medical_records}. We first introduce the dataset
 used for training the signatures and similarity functions, then assess the
embedding performance and retrieval performance of DDV on the real dataset,
and finally discuss how the different noise levels protect the signature and
effect retrieval performance. 

\paragraph{Dataset}
We use a real EHR dataset from a hospital that includes more than 283,000 patients with documented diagnostic histories.
A typical patient diagnostic history consists of a list of relevant ICD-9 diagnostic group codes for each visit as well as a summary ICD-9 problem list for the patient. 
We emphasize here that the difference between \emph{ICD-9 group codes} and \emph{ICD-9 codes} lies in the fact that \emph{group codes} consist of 3-letter representations of original ICD-9 codes. For example, a `\emph{250.00}' is the ICD-9 code for \emph{type-II diabetes (T2DM) w/o specified complications}, while '\emph{250}' is the general ICD-9 group code for \emph{type-II diabetic-related diseases}, including T2DM with and without various specified complications. The group codes are typically used in feature representation of patients, while the ICD-9 codes themselves provide greater granularity in terms of querying during cohort-selection.

\paragraph {Embedding Performance of Patient-Level Signatures}
In the representation task, we try to capture medical history information in the EHR database by finding a matrix representation $f_{\theta}(g_{\rho}(\mathcal{X}))= \mathbf{X'}$ of the visit histories of each patient over the entire dataset tensor, $\mathcal{X}$. Here, the vector representation $\mathbf{x'}_i \in \mathbf{X'}$ should ideally capture the sparsity features as well as the local temporal informations from the diagnostic history matrix $\mathbf{x}_i \in \mathcal{X}$ for each patient. We therefore evaluate the recovery performance of our representation across each visit and at each timestep for all patients. 

In total, there are 1041 group codes in our dataset, so training inputs take
on $\mathbf{x}_i \in \mathbb{R}^{1041}$ one-hot vector format, with $1$'s denoting
group codes present for the visit. On average, each patient visit averages $3.69$ group codes out of
the possible $1041$. Thus, we give greater weight to sparse, positive labels across
all visits during training. Otherwise, an embedding which recovers all $0$ entries over $1041$
group codes for each visit can still achieve $0.99$ accuracy due to the
prominence of negative codes.
 
We evaluate the effectiveness of the patient-level
autoencoder in two ways (1) given the de-noised signature vector for each patient, 
evaluate the micro-averaged precision and recall across all visits
over 5-fold validation and (2) after addition of Gaussian noise $\epsilon$ protection to each signature vector, re-train the decoder to recover 
the original $\mathcal{X}$ and evaluate performance over 5-fold validation. 
When a consumer purchases a substantial amount of medical records, the individual can theoretically train a 
decoder to recover the original medical dataset from the publicly available signature vectors. 
By adding Gaussian noise to the signatures, we can significantly limit the ability of an accurate decoder to be trained
and generalized to the rest of the dataset. The noise-addition step thus serves as a privacy-preserving procedure over publicly available signature vectors. By comparing performance of the recovery processes in (1) and (2),
we can gauge the upper limit in the level of granularity that a decoder is capable of recovering, given various levels of noise during the privacy-preserving process. Thus, (2) is included in our evaluation procedure as a measure of "decoder protection" which can be tracked in real time during deployment.


\begin{figure}[t!]
\centering
\includegraphics[width=0.48\textwidth]{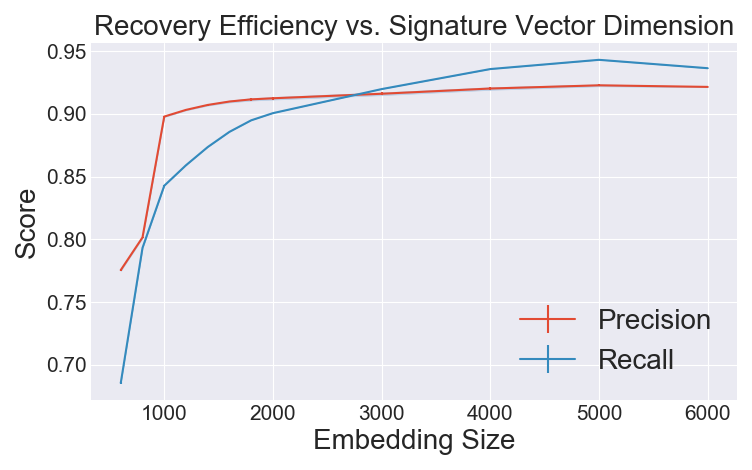}
\caption{Recovery efficiency of autoencoder framework using various signature vector dimension sizes.}
\label{fig:recovery}
\end{figure}

From Figure~\ref{fig:recovery}, we see that the embedding quality, as measured by recovery precision and recall over $\mathcal{X}$ from $f^{-1}(f_{\theta}(g_{\rho} (\mathcal(X))))$, improves greatly with increasing intermediate signature vector representation size up to saturation point of around 2000 dimensions. At 6000 dimensions, signature vectors were able to capture over $90\%$ of the sparse temporal features from the original data. This is unsurprising, as each visit vector from the original dataset was 1041 in dimension, and the average patient had $6.65$ visits in their diagnostic histories. It is notable that at 2000 feature dimensions, recovery over $90\%$ precision and recall was still achievable. 

Figure \ref{fig:recovery} further illustrates the saturation of recovery efficiency with increase in signature vector sizes. Here we see that the shoulder of the performance curve occurs at 1000 dimensions, where increases in recovery efficiency decreases dramatically with additional embedding dimensions. For the purposes of scalable feature representation and retrieval performance, we want to use an embedding size which provides adequate compression of feature space from the original data without drastically sacrificing embedding accuracy. We thus recommend using 1000 dimensional embedding representation, as it provides over 7x compression of the original medical data while preserving majority of the sparsity and temporal properties for downstream tasks.  

\begin{figure}[t!]
\small
\centering
\includegraphics[width=0.48\textwidth]{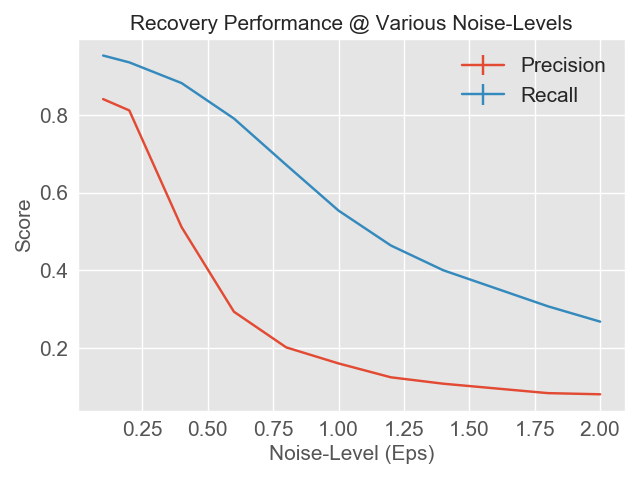}
\caption{Decoder recovery performance at various noise levels on 1000 dimensional vector signatures.}
\label{fig:eps_recovery}
\end{figure}

\paragraph{Decoder Protection}
When considering embedding performance a key question to consider is: can we maintain the main
information in the signature vector after adding protection noise? To
answer this question we use the original decoder (jointly trained with the
encoder) to recover data matrices from noise protected signature vectors. 
In Figure~\ref{fig:eps_recovery}, we demonstrate the robustness of decoder recovery to privacy-preserving Gaussian noise. Again, we witness an inflection point in the ability of the decoder to learn an accurate recovery of the original data under various noise levels. Here, we center the Gaussian noise at mean of $\mu = 0.0$ and increase the standard deviation of noise, $\sigma = \epsilon$ where $\epsilon \in \{0.10, 0.20, 0.40, 0.60, 0.80, 1.00, 1.20, 1.40, 1.80, 2.00\}$. 
At each noise level $\epsilon$, we construct a noise vector $\nu \in \mathbb{R}^d$, where $d$ is dimension of signature vector, and each component of $\nu$ consists of a random scalar drawn from $\mathcal{N}(\mu, \epsilon)$. For each vector, we construct a different noise vector, drawn from the same distribution under $\mathcal{N}(\mu, \epsilon)$. This noise vector then gets added to the original signature vector $\mathbf{X}_i$.

As seen in Table \ref{table:table_eps_recovery}, the steepest drop in recovery performance occurred between $ 0.40 \leq \epsilon \leq 0.60$, where precision decreased from above $80\%$ to below $50\%$. Further increases in noise levels facilitate the exponential decrease in decoder recovery, yet it is interesting to note that the exponential decay of decoder performance did not occur until noise level reached a substantial level. This is most likely due to the inherent robustness of the decoder to noise at the input level. 
Another interesting pattern to note is that added noise induced more false-positive predictions by the decoder, which is indicated by the increasing between precision and recall scores with increasing noise levels.

Given the retrieval performance under noise protection, one key question still remains: can we prevent users from recovering the decoder when they have accumulated a reasonable amount of protected signatures? 
We answer this question by measuring decoder parameter changes during the re-training procedure 
by first vectorizing the weight layers in the decoders and then taking the Euclidean distance between the de-noised parameters $\theta_0$ and re-trained parameters $\theta_{\epsilon}$ at each noise level. 
Table~\ref{table:thetas_eps}, examines these parametric differences between 
the original decoder and the recovered one. 
One clear trend is that as the magnitude of noise increases, differences in parameter decrease during re-training. 
With increasing noise, the decoder is less capable of learning correct parameters due to the increasing randomness in the added noise. 
It is important to note that while increase in noise level can provide privacy over the signature vector signatures, the optimal level of noise must be fine tuned against retrieval efficiency. In the following section, we demonstrate the trade-off between the protectiveness of high $\epsilon$ noise-levels and the usability of the transformed signature vectors for retrieval. 

\begin{table}[t!]
\small
\centering
\caption{Recovery performance using 1000 dimensional signature under various noise levels.}
\begin{tabular}{c||c|c}
\hline
\textbf{$\epsilon$} & Precision & Recall  \\ \hline
0.10 & 0.841 $(\pm8.8\times 10^{-4})$ & 0.953 $(\pm3.1\times 10^{-5})$ \\
0.20 & 0.811 $(\pm2.7\times 10^{-5})$ & 0.935 $(\pm5.0\times 10^{-6})$ \\
0.40 & 0.511 $(\pm2.9\times 10^{-4})$ & 0.882 $(\pm2.0\times 10^{-5})$ \\
0.60 & 0.293 $(\pm2.8\times 10^{-4})$ & 0.791 $(\pm5.1\times 10^{-5})$ \\
0.80 & 0.202 $(\pm5.3\times 10^{-5})$ & 0.672 $(\pm1.8\times 10^{-5})$ \\
1.00 & 0.160 $(\pm6.6\times 10^{-6})$ & 0.553 $(\pm2.8\times 10^{-5})$ \\
1.20 & 0.124 $(\pm2.3\times 10^{-5})$ & 0.464 $(\pm7.5\times 10^{-5})$ \\
1.40 & 0.108 $(\pm6.8\times 10^{-6})$ & 0.400 $(\pm7.7\times 10^{-5})$ \\
1.80 & 0.084 $(\pm9.3\times 10^{-7})$ & 0.300 $(\pm2.2\times 10^{-6})$ \\
2.00 & 0.080 $(\pm1.3\times 10^{-6})$ & 0.268 $(\pm2.5\times 10^{-6})$ \\
\hline
\end{tabular}
\label{table:table_eps_recovery}
\end{table}

\begin{table}[t!]
\small
\centering
\caption{Changes in learned decoder parameters with increasing noise.}
\begin{tabular}{c||c|c}
\hline
\textbf{$\epsilon$} & \textbf{$||\theta_{\epsilon} - \theta_0||_2^2$} & \textbf{$||\theta_{\epsilon} - \theta_0||_2^2 / ||\theta_{0}||_2^2$}\\ \hline
0.10 & 91.96 & 0.402 \\
0.20 & 90.56 & 0.395 \\
0.40 & 85.94 & 0.376 \\
0.60 & 85.75 & 0.375 \\
0.80 & 87.51 & 0.382 \\
1.00 & 90.16 & 0.393 \\
1.20& 83.70 & 0.366 \\
1.40& 81.64 & 0.357 \\
1.80& 81.53 & 0.355 \\
2.00& 81.64 & 0.360 \\
\hline
\end{tabular}
\label{table:thetas_eps}
\end{table}

\paragraph{Retrieval Performance}
From the standpoint of the consumer, a typical query order on a medical database include
selecting for patients with specific ICD-9 codes for downstream cohort studies. 
For this reason, we formulate the retrieval task as searching for top \emph{N} patients, given a specific cohort.  
We constructed 10 mutually exclusive cohorts, distinguished by ICD-9 codes.  
Table \ref{table:cohorts} summarizes the number of available patients for each cohort. 
In total, there were 58,489 patients in this subset of the EHR database. 
Each patient has a unique label in one-hot vector format, $\mathbf{y} \in \mathbb{R}^{10}$.
The retrieval task is formulated as follows: \textit{given a cohort of interest, $\mathbf{y}$, find the top N patients that belonging to this cohort}. 

As mentioned in the previous sections, a similarity matrix $M_t$ is trained using multi-task metric learning~\cite{parameswaran2010large} for each task $\mathbf{y}_t \in \{\mathbf{y}_1, ..., \mathbf{y}_{10}\}$. 
Then, a characteristic query vector $\mathbf{q}_t$ is obtained for each task by taking the component-wise average of signature vectors across the cohort. 
For example, the congestive heart failure (CHF) cohort contains $1,456$ unique patients, each represented by a signature vector $\mathbf{x'}_i \in \mathbb{R}^{d}$. 
The cohort can thus be represented by a $1,456 \times d$ matrix, $\mathbf{X'}_{\text{CHF}}$. 
To perform the retrieval task, we take the characteristic query vector to be $\mathbf{q} = \frac{1}{m} \sum_i^{m} \mathbf{x'}_i$, where $m=1,456$.  
Once this query vector is obtained, $M_t$ can be used to calculate the distance for each patient in the cohort, where $d_t(\mathbf{q}_t, \mathbf{x'}_i) = \sqrt{(\mathbf{q}_t - \mathbf{x'}_i)M_t(\mathbf{q}_t - \mathbf{x'}_i)}$. 
We then rank the top $N$ ``closest'' patients to the target query by Mahalanobis distance and use them for retrieval evaluation. 

We evaluate the performance of each retrieval task by calculating the \emph{precision@N} score for $N \in \{10, 50, 100, 250, 500\}$. 
By comparing precision at various retrieval sizes ($N$), we can deduce the scalability of various queries, given the metrics $M_t \in \mathcal{M}$ learned by ~\cite{parameswaran2010large}. Table \ref{table:retrieval} demonstrates these differences in scalability, as we can see that for tasks such as MCI, CAD and CHF, retrieval precision decreases dramatically after 100 cohorts, while for tasks such as T2DM, lung cancer and breast cancer retrieval, precision remains near $100\%$ even at much larger retrieval sizes. By contrast, the actual largest margin nearest neighbors (MT-LMNN) performance, as shown in Table \ref{table:mtlmnn}, indicates that the classification error remain consistently below $7\%$ across all tasks when classification is done in a pair-wise manner between candidate patients. The cross-cohort discrepancy between query-based retrieval and MT-LMNN performance highlights the importance of query design for accurate retrieval under our framework. 

For example, in the case of T2DM, predictable patterns of comorbidity often ensue the progression of the disease. End-stage renal diseases (ESRD), ophthalmic and neurosensory complications inevitably co-occur with T2DM during the longitudinal development of symptomatology. 
On the other hand, CHF does not follow a singular, predictable pattern of development. In fact, CHF itself can be decomposed into several subtypes of progression, all of which result from vastly different compensatory pathways. Since the retrieval task relies on a single query representation $\mathbf{q}_{t}$ per cohort, techniques such as signature-vector averaging and singular vector selection may not adequately capture the variability within certain cohorts. Future studies may focus on designing effective ways to produce query vector representations which can cover such variability within disease phenotypes.

\begin{table}[t!]
\small
\centering
\caption{Summary of cohorts for retrieval tasks.}
\begin{tabular}{c|c|c|c}
\hline
ICD-9 Code & Description &Raw Count &Exclusive \\ \hline\hline
250.00 & Type-II DM & 39,699 & 13,381  \\
427.31 & Arrythmias & 21,072 & 4,734  \\
311 & Depression & 29,978 & 9,863  \\
272.4 & Hyperlipidemia & 48,302 & 12,500  \\
414.00 & CAD & 22,936 &  3,249 \\
244.9 & Hypothyroidism & 22,416 & 6,167 \\
162.9 & Lung Cancer & 5,397 & 2,927 \\
428.0 & CHF & 13,516 & 1,456 \\
174.9 & Breast Cancer & 8,379 & 3,672 \\
331.83 & MCI & 2,473 & 540 \\
\hline
Total & & & 58,489 \\
\hline
\end{tabular}
\label{table:cohorts}
\end{table}

\begin{table}[t!]
\small
\centering
\caption{Multi-task LMNN classification error rates.}
\begin{tabular} { l | c | c  } 
\hline
Task & Validation & Testing \\ 
\hline\hline
T2DM & 2.750 & 2.621 \\ 
Arrythmia & 4.912 & 6.604 \\ 
Depression & 2.292 & 3.564 \\ 
Hyperlipidemia & 6.549 & 6.918 \\ 
CAD & 3.274  & 4.298 \\ 
Hypothyroidism & 3.602 & 3.826 \\ 
Lung Cancer & 0.655 & 0.786 \\ 
CHF & 3.274 & 3.878 \\ 
Breast Cancer & 1.834 & 1.992 \\ 
MCI & 1.965 & 2.306 \\ 
\hline
\end{tabular}
\label{table:mtlmnn}
\end{table}

\begin{table}[t!]
\centering
\caption{Retrieval performance of learned cohort-specific metrics using 1000 dimensional de-noised signatures. Top performing queries include Type-II Diabetes Mellitus (T2DM), lung and breast cancers, all of which achieved $\geq$ 99\% precision at various retrieval sizes.}
\begin{tabular}{c|c|c|c|c|c}
\hline
Cohort & Prc@10 & Prc@50 & Prc@100 & Prc@250 & Prc@500 \\ 
\hline\hline            
T2DM& 1.000  & 1.000 & 1.000 & 1.000 & 1.000 \\                                     
Arrythmia& 1.000 & 1.000 & 1.000 & 0.684 & 0.408 \\
Breast Cancer & 1.000 & 1.000 & 1.000 & 0.996 & 0.988 \\
CAD& 1.000 & 0.935 & 0.581 & 0.352 & 0.194 \\
CHF& 1.000 & 0.940 & 0.466 & 0.200 & 0.106 \\
Depression& 1.000 & 1.000 & 1.000 & 1.000 & 0.686 \\
Hyperlipidemia& 1.000 & 1.000 & 0.794 & 0.784 & 0.608 \\
Hypothyroid& 1.000 & 0.981  & 0.979  & 0.624 & 0.408 \\
Lung Cancer& 1.000& 1.000 & 1.000 & 1.000 & 1.000 \\
MCI&  1.000 & 0.781  & 0.443  & 0.184 & 0.100 \\
\hline
\end{tabular}
\label{table:retrieval}
\end{table}

\paragraph{Effect of Noise on Retrieval}
As mentioned previously, privacy-preserving noise produces a trade-off in performance between recovery and retrieval:
addition of high noise may decrease decoder recovery, thereby increasing privacy, yet it may also decrease retrieval performance.
We examine the effect of noise on retrieval performance by re-learning similarity matrices $M_t \in \mathcal{M}$ for each individual task while using the increasingly noisy versions of signature vectors. 
Retrieval performance was again evaluated using \emph{precision@N} at each noise level $\epsilon \in \{0.0, 0.2, 0.4, 0.6, 0.8, 1.0\}$ across various retrieval tasks. This time, we included only the top performing retrieval tasks, including: T2DM, lung cancer, breast cancer, depression and arrhythmia. These tasks possess the highest performing queries from the de-noised retrieval setting. 

\begin{figure}
    \centering
    \begin{subfigure}[t]{0.24\textwidth}
        \centering
        \includegraphics[width=\linewidth]{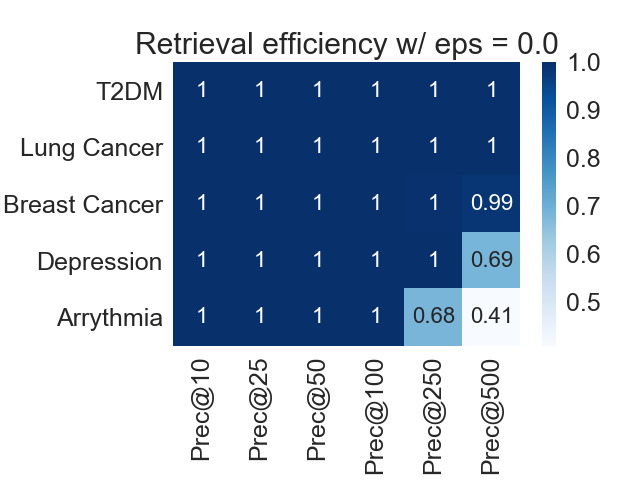} 
        \caption{$\epsilon = 0.0$} \label{fig:heatN0}
    \end{subfigure}
    \hfill
    \begin{subfigure}[t]{0.24\textwidth}
        \centering
        \includegraphics[width=\linewidth]{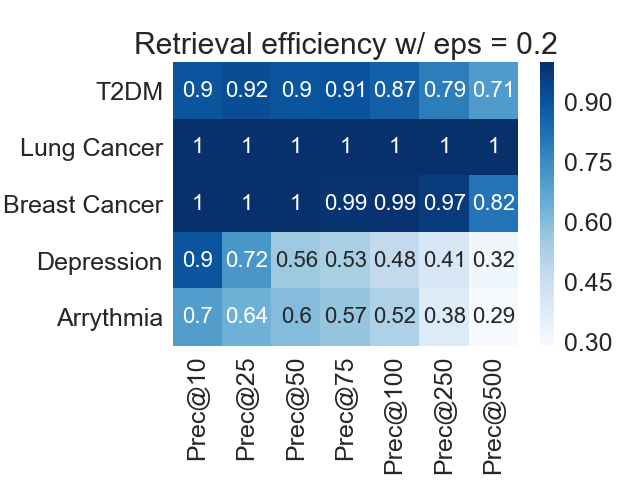} 
        \caption{$\epsilon = 0.2$} \label{fig:heatN02}
    \end{subfigure}

    \vspace{.1cm}
    \begin{subfigure}[t]{0.24\textwidth}
    \centering
        \includegraphics[width=\linewidth]{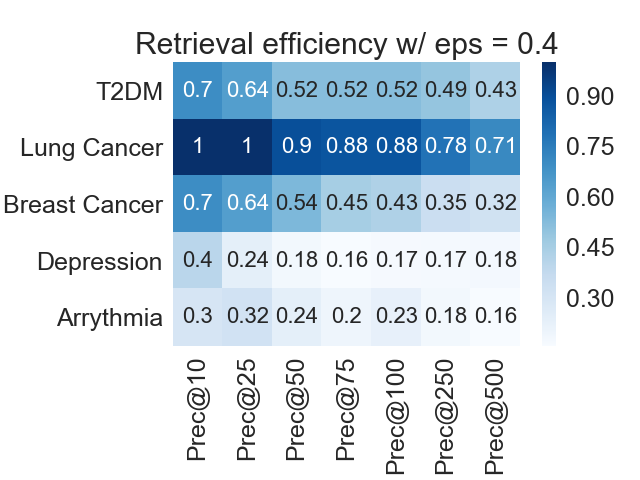} 
        \caption{$\epsilon = 0.4$} \label{fig:heatN04}
    \end{subfigure}
    \hfill
        \begin{subfigure}[t]{0.24\textwidth}
        \centering
        \includegraphics[width=\linewidth]{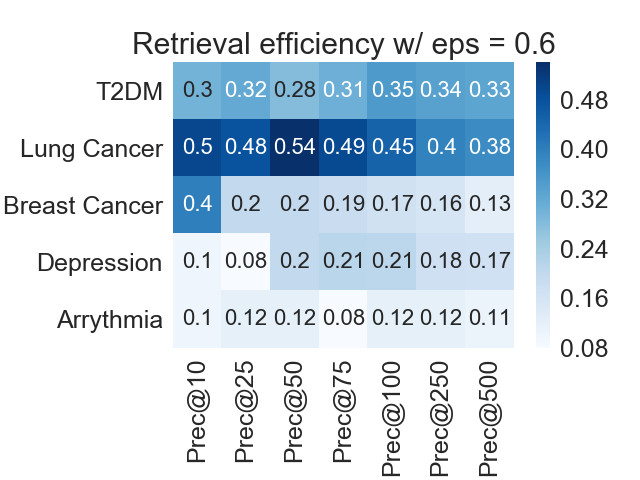} 
        \caption{$\epsilon = 0.6$} \label{fig:heatN06}
    \end{subfigure}
    \vspace{.1cm}
        \begin{subfigure}[t]{0.24\textwidth}
        \centering
        \includegraphics[width=\linewidth]{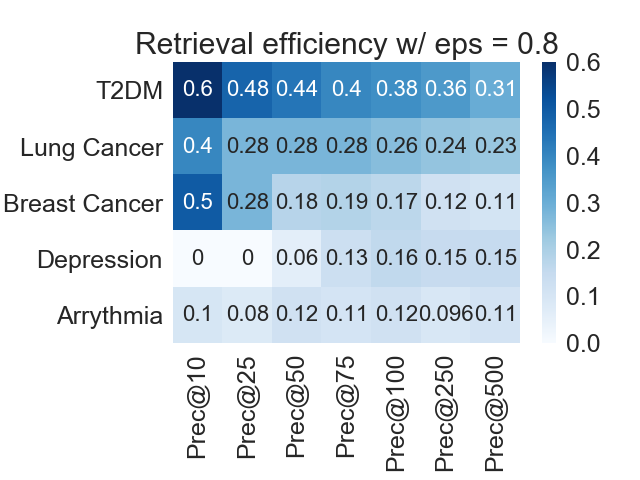} 
        \caption{$\epsilon = 0.8$} \label{fig:heatN08}
    \end{subfigure}
    \hfill
    \begin{subfigure}[t]{0.24\textwidth}
        \centering
        \includegraphics[width=\linewidth]{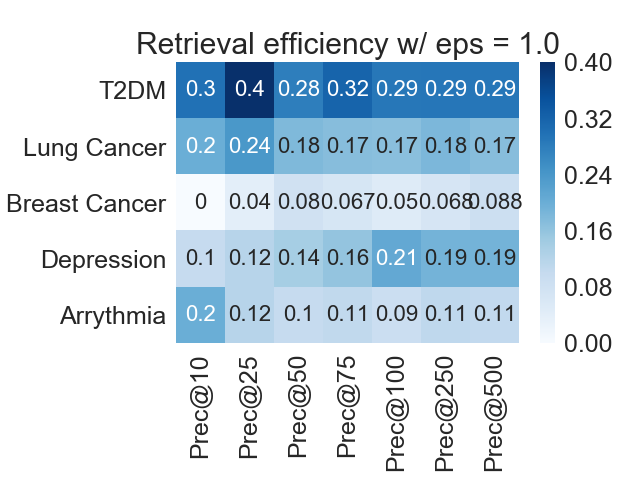} 
        \caption{$\epsilon = 1.0$} \label{fig:heatN1}
    \end{subfigure}

    \caption{Heatmap displays of retrieval efficiency across various retrieval demands with increasing noise levels.}
\end{figure}

As shown in Figures \ref{fig:heatN02} - \ref{fig:heatN1}, precision scores drop dramatically with increasing noise levels. With $\epsilon > 0.6$, the recovery behavior becomes more erratic and may not follow the monotonically decreasing trend as seen with the typical case. Figures \ref{fig:heatN02} and \ref{fig:heatN04} demonstrate the noise-levels that are capable of producing accurate retrievals for majority of the tasks. When comparing with the privacy-preserving performance shown in Figure \ref{fig:eps_recovery}, we see that the optimal noise level for 1000 dimension signatures is $\epsilon = 0.4$, as this level of noise disrupts decoder learning enough to limit the consumer's capacity to recover the original dataset while maintaining precision during retrieval. We notice that in both recovery and retrieval, noise levels beyond $\epsilon = 0.6$ leads to erratic behavior. In the case of recovery, decoder performance decreases exponentially with increasing noise above $0.6$, but retrieval performance suffers greatly in response. 

\section{Conclusion}
In this paper, we studied the problem of \emph{distributed data vending},
which allows personal data to be securely exchanged on a blockchain. We
proposed a framework for distributed data vending through a combination of
data embedding and similarity learning, to tackle the dilemma between the
effectiveness of data retrieval, and the leakage from indexing the data.
Distributed data vending can transform domains such as healthcare by
encouraging data distribution from owners and its aggregation. We illustrated
our framework through a practical scenario of distributing electronic medical
records on a blockchain. We conducted extensive empirical results to
demonstrate the effectiveness of our framework.

\section*{Acknowledgment}
\noindent This research is supported in part by VeChain Foundation.
We thank
Chi Jin from Michigan State for discussions during the work. 

\bibliographystyle{abbrv}
\bibliography{ref}

\begin{thebibliography}{10}

\bibitem{azaria2016medrec}
A.~Azaria, A.~Ekblaw, T.~Vieira, and A.~Lippman.
\newblock Medrec: Using blockchain for medical data access and permission
  management.
\newblock In {\em Open and Big Data (OBD), International Conference on}, pages
  25--30. IEEE, 2016.

\bibitem{baxendale2016can}
G.~Baxendale.
\newblock Can blockchain revolutionise eprs?
\newblock {\em ITNow}, 58(1):38--39, 2016.

\bibitem{brodersen2016applying}
C.~Brodersen, B.~Kalis, C.~Leong, E.~Mitchell, E.~Pupo, and A.~Truscott.
\newblock Applying blockchain technology to medicine traceability, 2016.

\bibitem{brown2016patient}
S.-A. Brown.
\newblock Patient similarity: Emerging concepts in systems and precision
  medicine.
\newblock {\em Frontiers in physiology}, 7:561, 2016.

\bibitem{cui2005content}
Y.~Cui, N.~Shivakumar, A.~Carobus, D.~Jindal, and S.~Lawrence.
\newblock Content-targeted advertising using collected user behavior data,
  Jan.~27 2005.
\newblock US Patent App. 10/649,585.

\bibitem{ekblaw2016medrec}
A.~Ekblaw and A.~Azaria.
\newblock Medrec: Medical data management on the blockchain.
\newblock {\em Ariel}, 1(10):7, 2016.

\bibitem{esposito2014health}
L.~Esposito.
\newblock How to get access to your hospital records, 2014.

\bibitem{greene2017targeting}
J.~A. Greene and W.~V. Padula.
\newblock Targeting unconscionable prescription-drug prices—maryland’s
  anti--price-gouging law.
\newblock {\em New England Journal of Medicine}, 377(2):101--103, 2017.

\bibitem{iansiti2017truth}
M.~Iansiti and K.~R. Lakhani.
\newblock The truth about blockchain.
\newblock {\em Harvard Business Review}, 95(1):118--127, 2017.

\bibitem{jolliffe1986principal}
I.~T. Jolliffe.
\newblock Principal component analysis and factor analysis.
\newblock In {\em Principal component analysis}, pages 115--128. Springer,
  1986.

\bibitem{karatzoglou2010multiverse}
A.~Karatzoglou, X.~Amatriain, L.~Baltrunas, and N.~Oliver.
\newblock Multiverse recommendation: n-dimensional tensor factorization for
  context-aware collaborative filtering.
\newblock In {\em Proceedings of the fourth ACM conference on Recommender
  systems}, pages 79--86. ACM, 2010.

\bibitem{koren2009matrix}
Y.~Koren, R.~Bell, and C.~Volinsky.
\newblock Matrix factorization techniques for recommender systems.
\newblock {\em Computer}, 42(8), 2009.

\bibitem{kuo2017blockchain}
T.-T. Kuo, H.-E. Kim, and L.~Ohno-Machado.
\newblock Blockchain distributed ledger technologies for biomedical and health
  care applications.
\newblock {\em Journal of the American Medical Informatics Association},
  24(6):1211--1220, 2017.

\bibitem{kuo2018modelchain}
T.-T. Kuo and L.~Ohno-Machado.
\newblock Modelchain: Decentralized privacy-preserving healthcare predictive
  modeling framework on private blockchain networks.
\newblock {\em arXiv preprint arXiv:1802.01746}, 2018.

\bibitem{linden2003amazon}
G.~Linden, B.~Smith, and J.~York.
\newblock Amazon. com recommendations: Item-to-item collaborative filtering.
\newblock {\em IEEE Internet computing}, 7(1):76--80, 2003.

\bibitem{livingston2018health}
S.~Livingston.
\newblock Health insurers' proposed 2018 rate hikes are early 'warning signs',
  2018.

\bibitem{luong2015multi}
M.-T. Luong, Q.~V. Le, I.~Sutskever, O.~Vinyals, and L.~Kaiser.
\newblock Multi-task sequence to sequence learning.
\newblock {\em arXiv preprint arXiv:1511.06114}, 2015.

\bibitem{mettler2016blockchain}
M.~Mettler.
\newblock Blockchain technology in healthcare: The revolution starts here.
\newblock In {\em e-Health Networking, Applications and Services (Healthcom),
  2016 IEEE 18th International Conference on}, pages 1--3. IEEE, 2016.

\bibitem{mika1999fisher}
S.~Mika, G.~Ratsch, J.~Weston, B.~Scholkopf, and K.-R. Mullers.
\newblock Fisher discriminant analysis with kernels.
\newblock In {\em Neural networks for signal processing IX, 1999. Proceedings
  of the 1999 IEEE signal processing society workshop.}, pages 41--48. Ieee,
  1999.

\bibitem{nakamoto2008bitcoin}
S.~Nakamoto.
\newblock Bitcoin: A peer-to-peer electronic cash system.
\newblock 2008.

\bibitem{us2003hippa}
U.~D. of~Health, H.~Services, et~al.
\newblock Summary of the hipaa privacy rule.
\newblock {\em Washington, DC: Author. Retrieved December}, 2:2007, 2003.

\bibitem{parameswaran2010large}
S.~Parameswaran and K.~Q. Weinberger.
\newblock Large margin multi-task metric learning.
\newblock In {\em Advances in neural information processing systems}, pages
  1867--1875, 2010.

\bibitem{peterson2016blockchain}
K.~Peterson, R.~Deeduvanu, P.~Kanjamala, and K.~Boles.
\newblock A blockchain-based approach to health information exchange networks.
\newblock In {\em Proc. NIST Workshop Blockchain Healthcare}, volume~1, pages
  1--10, 2016.

\bibitem{sutskever2014sequence}
I.~Sutskever, O.~Vinyals, and Q.~V. Le.
\newblock Sequence to sequence learning with neural networks.
\newblock In {\em Advances in neural information processing systems}, pages
  3104--3112, 2014.

\bibitem{livingston2016health}
P.~Taylor.
\newblock Applying blockchain technology to medicine traceability, 2016.

\bibitem{topol2016money}
E.~J. Topol.
\newblock Money back guarantees for non-reproducible results?, 2016.

\bibitem{tschorsch2016bitcoin}
F.~Tschorsch and B.~Scheuermann.
\newblock Bitcoin and beyond: A technical survey on decentralized digital
  currencies.
\newblock {\em IEEE Communications Surveys \& Tutorials}, 18(3):2084--2123,
  2016.

\bibitem{vechain2018web}
Vechain.
\newblock Vechain, 2018.

\bibitem{velimeneti2016data}
S.~Velimeneti.
\newblock Data migration from legacy systems to modern database.
\newblock 2016.

\bibitem{vincent2008extracting}
P.~Vincent, H.~Larochelle, Y.~Bengio, and P.-A. Manzagol.
\newblock Extracting and composing robust features with denoising autoencoders.
\newblock In {\em Proceedings of the 25th international conference on Machine
  learning}, pages 1096--1103. ACM, 2008.

\bibitem{wood2014ethereum}
G.~Wood.
\newblock Ethereum: A secure decentralised generalised transaction ledger.
\newblock {\em Ethereum Project Yellow Paper}, 151:1--32, 2014.

\bibitem{yang2006distance}
L.~Yang and R.~Jin.
\newblock Distance metric learning: A comprehensive survey.
\newblock {\em Michigan State Universiy}, 2(2), 2006.

\bibitem{zhou2011malsar}
J.~Zhou, J.~Chen, and J.~Ye.
\newblock Malsar: Multi-task learning via structural regularization.
\newblock {\em Arizona State University}, 2011.

\bibitem{zhou2013modeling}
J.~Zhou, J.~Liu, V.~A. Narayan, J.~Ye, A.~D.~N. Initiative, et~al.
\newblock Modeling disease progression via multi-task learning.
\newblock {\em NeuroImage}, 78:233--248, 2013.

\bibitem{zhou2013feafiner}
J.~Zhou, Z.~Lu, J.~Sun, L.~Yuan, F.~Wang, and J.~Ye.
\newblock Feafiner: biomarker identification from medical data through feature
  generalization and selection.
\newblock In {\em Proceedings of the 19th ACM SIGKDD international conference
  on Knowledge discovery and data mining}, pages 1034--1042. ACM, 2013.

\bibitem{zhou2014micro}
J.~Zhou, F.~Wang, J.~Hu, and J.~Ye.
\newblock From micro to macro: data driven phenotyping by densification of
  longitudinal electronic medical records.
\newblock In {\em Proceedings of the 20th ACM SIGKDD international conference
  on Knowledge discovery and data mining}, pages 135--144. ACM, 2014.

\bibitem{zhu2018edgechain}
H.~Zhu, C.~Huang, and J.~Zhou.
\newblock Edgechain: Blockchain-based multi-vendor mobile edge application
  placement.
\newblock {\em arXiv preprint arXiv:1801.04035}, 2018.

\end{thebibliography}

\end{document}